\documentclass{article}

\usepackage{arxiv}

\usepackage[utf8]{inputenc} 
\usepackage[T1]{fontenc}    
\usepackage{hyperref}       
\usepackage{url}            
\usepackage{booktabs}       
\usepackage{amsfonts,amsmath,amssymb}       
\usepackage{nicefrac}       
\usepackage{microtype}      
\usepackage{amsthm,wrapfig,pgf,tikz,pgfplots}
\usepackage{graphicx}
\usepackage{dcolumn}
\usepackage{bm}
\usepackage{caption,subcaption}

\pgfplotsset{compat=1.14}
\usetikzlibrary{arrows}

\title{The Hybrid Quasiparticles in Organic- Semiconductor Quantum Dot System}

\author{
  David W. Facemyer\\
  Department of Physics\\
  Marshall University\\
  Huntington  WV 25755 \\
  \texttt{facemyer3@marshall.edu} \\
   \And
  Nguyen Que Huong\\
  Department of Physics\\
  Marshall University\\
  Huntington  WV 25755 \\
  \texttt{nguyenh@marshall.edu} \\
}

\begin{document}
\maketitle

\begin{abstract}
In this work electronic structures and optical properties of organic-inorganic exciton and polaritons in two-dimensional heterostructures combining both organic and semiconductor materials are studied. In those systems, Wannier-Frenkel hybrid exciton has unique and interesting properties that can improve the efficiency of optical materials. When an organic-semiconductor combined heterostructure is illuminated by high-intensity electromagnetic radiation with the frequency of the photons at or near the resonance frequency of the Wannier-Frenkel exciton, we obtain a macroscopically occupied system of hybrid polaritons. We will theoretically determine electronic structure, energy and dispersion relation of the hybrid excitons and polaritons. By analyzing the parameters of the systems and the interactions between the Wannier and Frenkel excitons and the photons, we then  discuss the conditions for hybridization. 
\end{abstract}

\keywords{hybrid exciton \and quantum dot \and polariton}

\section{\label{Introduction}Introduction}
Recent development of technology opens a new opportunity of incorporating organic and semiconductor materials in combined low-dimensional heterostructure systems. Quantum size effects of the motion of electrons and holes in low-dimensional heterostructures lead to many interesting optical and structural phenomena. 

In semiconductors, Wannier-Mott excitons have large exciton radii and weak oscillator strengths, large exciton-exciton interaction and low saturation exciton density. In organic materials, in contrast, small-radius Frenkel excitons have large oscillator strength, small exciton-exciton interaction and very large saturation density. When a semiconductor is placed next to an organic material in a combined organic-inorganic heterostructure configuration, the Wannier exciton and the Frenkel exciton interact with each other and a special kind of hybridization state, the hybrid exciton,  is formed at a junction or boundary between semiconductor and organic materials due to the dipole-dipole interaction. Hybrid exciton formation in different systems has been studied in different geometrical configurations \cite{engelmann,AG03,AG94,AG95,QH1,QH2,QH3,QH4} and has been shown to possess many desirable properties such as large exciton Bohr radius, large oscillator strength and  high optical nonlinearities.

Large exciton radius especially favors polariton-polariton interaction and increases the operational temperature. The coupling of the two states is enhanced if the respective energy levels are in or near resonance. The hybridization of exciton state suggests using the organic component to dramatically enhance the operational temperature and the optical nonlinearities of inorganic structures. The observation of cavity mediated hybridization of GaAs and J-aggregate exciton in the strong coupling regime under electrical injection of carriers and the polariton lasing up to 200 K under non-resonant optical pumping \cite{Paschos} open the doors for successful realization of an inorganic-organic hybrid polariton laser with stable high polariton operational temperature and highly efficient electrical injection in the very near future. Recent work \cite{Jayaprakash} shows that, by electrically injecting electron–hole pairs into the inorganic quantum-well system, a transfer of energy between cavities is possible, populating organic-exciton polaritons and exploring a new way to facilitate energy transfer between semiconductor materials.

In this work we theoretically study the generalized model of hybrid excitons and hybrid polaritons in different geometric configurations.  In addition to the effects certain parameters have on a system, numerical models of the dispersion relation of each system's hybrid polariton, all within an \textit{ab initio} framework, are analyzed in order to discuss specific environments that promote the most favorable attributes of the system. From this theoretical information and the use of empirical data, we are then able to suggest components of heterostructures that foster ideal systems tailored to need and application.

\section{\label{Hybridization}Wannier-Frenkel Exciton Hybridization}
Let us consider a structure consisting of a semiconductor component and an organic component. The theory could be applied for different semiconductor-organic heterostructures. One of the examples is a semiconductor quantum dot coated by an organic shell. The total Hamiltonian of the system of interacting Wannier excitons and Frenkel excitons will be taken as follows: 
\begin{equation}
\label{InteractionHamiltonian}
H_{\alpha}=H_{F}+H_{W}+H_{F-W}=\sum_{\mathbf{k}}{\left[E_{\mathbf{k}}^{F}\hat a_{\mathbf{k}}^{\dagger}\hat a_{\mathbf{k}}+E_{\mathbf{k}}^{W}\hat b_{\mathbf{k}}^{\dagger}\hat b_{\mathbf{k}}+\Gamma_{\mathbf{k}}(\hat a_{\mathbf{k}}^{\dagger}\hat b_{\mathbf{k}}+\hat b_{\mathbf{k}}^{\dagger}\hat a_{\mathbf{k}})\right]},
\end{equation}
where $ H_F$ is the Frenkel exciton Hamiltonian,  $ H_W$ is the Wannier Mott exciton Hamiltonian and $H_{F-W}$ is the interaction Hamiltonian between the Wannier Mott excitons in the organic component  and the Frenkel excitons in the organic component with coupling constant $\Gamma_{\mathbf{k}}$. $E_{\mathbf{k}}^{W}$ and $E_{\mathbf{k}}^{F}$ are the excitation energies of Wannier excitons in the organic material and the Frenkel exciton in the organic medium, respectively. $\hat a_{\mathbf{k}}^{\dagger}(\hat a_{\mathbf{k}})$ are creation (annihilation) operators of Frenkel excitons, $\hat b_{\mathbf{k}}^{\dagger}( \hat b_{\mathbf{k}})$ are creation (annihilation) operators of Wannier excitons.

\begin{figure}[htbp]
\centering
\begin{subfigure}[b]{.5\textwidth}
  \centering
  \includegraphics[width=0.9\linewidth]{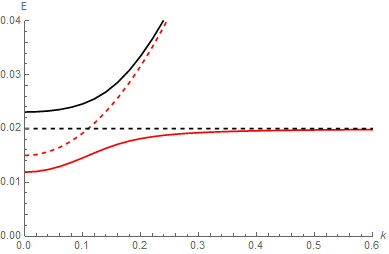}
  \caption{Dispersion relation for the hybrid exciton for $\Gamma_{\mathbf{k}} = 5 \, \mathrm{meV}.$}
  \label{hExcitonDispersionCurves}
\end{subfigure}%
\begin{subfigure}[b]{.5\textwidth}
  \centering
  \includegraphics[width=0.9\linewidth]{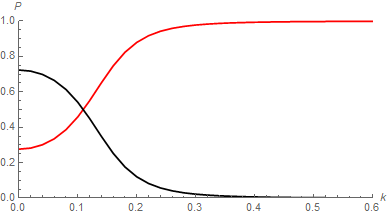}
  \caption{The $k$-dependence of the upper branch mixing coefficients.}
  \label{hExcitonUpperMixingCurves}
\end{subfigure}
\caption{Example of a hybrid exciton dispersion and branch mixing. The red and black curves reflect the red and black dotted curves of the pure states on the dispersion graph, respectively.}
\label{hybridexciton}
\end{figure}

Applying the unitary transformation we can write the system as a pure hybrid exciton system as follows
\begin{equation}
\label{HExcitonHamiltonian}
H_{\alpha}=\sum_{\mathbf{k}}{E_{\mathbf{k}}^{\alpha}\hat \alpha_{\mathbf{k}}^{\dagger}\hat \alpha_{\mathbf{k}}}.
\end{equation}
where $\hat \alpha_{\mathbf{k}}^{\dagger}(\hat \alpha_{\mathbf{k}})$ are creation (annihilation) operators of the Wannier-Frenkel exciton, and $E_{\mathbf{k}}^{\alpha}$ are the energies of the hybrid exciton. When the energies of Wannier and Frenkel excitons are comparable (i.e. when the energy separation between the Wannier-Mott and the Frenkel excitons is much less than the distance to other exciton bands) and the mixing state is assumed only between the two nearest bands, the basis of the hybridization is chosen such that when the Wannier-Mott exciton is in its excited state the Frenkel exciton is in its ground state and vice versa. We have the following:
\begin{equation}
|\Psi({\mathbf{k}})\rangle = u_l({\mathbf{k}})f^F(0)|\Psi^{W}_l({\mathbf{k}})\rangle + v_{l'}({\mathbf{k}})f^W(0)|\Psi^{F}_{l'}({\mathbf{k}})\rangle,
\end{equation}
with the branch-energy $E(\mathbf{k})$ of the hybrid state given as
\begin{equation}
\label{HExcitonEigenenergies}
E_{j}^\alpha(\mathbf{k}) = \frac{1}{2}\bigg( E_{\mathbf{k}}^F + E_{\mathbf{k}}^W \pm \sqrt{(E_{\mathbf{k}}^F - E_{\mathbf{k}}^W)^2 + 4\Gamma_{\mathbf{k}}^2} \bigg).
\end{equation}
The energy of the hybrid exciton as a function of $\mathbf{k}$, is shown in \textbf{Fig. \ref{hExcitonDispersionCurves}}, and the mixing coefficients for the upper hybrid exciton branch are shown in \textbf{Fig. \ref{hExcitonUpperMixingCurves}}, where the Frenkel exciton energy is considered independent of the wave vector $\mathbf{k}$.


The coupling constant (i.e. the hybridization parameter), $\Gamma_{\mathbf{k}}$, stands for the dipole-dipole interaction between excitons, taken as the product of the operator of the electric field $E(r)$ created at point $r$ in the organic medium by the excitons in quantum dots and the transition polarization operator  $P(r)$ of the Frenkel exciton at molecular site $r$ of the organic shell-medium,
\begin{equation}
\Gamma_{\mathbf{k}} = \langle F_u ,\mathbf{k}| U | W_v, \mathbf{k} \rangle,
\end{equation}
where, for the case of a semiconductor quantum dot with an organic shell heterostructure, $U$ is the interaction of the dipoles of the Frenkel exciton in the shell and the Wannier exciton in the semiconductor core. For the coated quantum dot heterostructure, let the dielectric constants for the dot core and shell be $\epsilon_c$ and $\epsilon_s$ respectively. For this we have,
\begin{equation}
\label{dipoleinteraction}
U = \frac{3 \epsilon_c}{2 \epsilon_s + \epsilon_c} \frac{(\vec{\mu}_F \cdot \vec{\mu}_W) - 3 (\vec{\mu}_W \cdot \hat{r}) (\vec{\mu}_F \cdot \hat{r})}{4 \pi \epsilon_0 {r}^3}.
\end{equation}
Then the coupling constant \textbf{(\ref{dipoleinteraction})} will be calculated for each of the excited states $u \neq v \in \{0,1\}$, which denotes respective ground states and first excited states.

In \cite{engelmann}, Engelmann and coauthors obtained the coupling constant as
\begin{equation}
\label{Engelmann}
\Gamma = - \frac{32 \mu_W \mu_F}{15 \epsilon} \sqrt{\frac{\pi R_{D}^3}{2 a_{B}^3 V_{\text{cell}} d_{0}^3}} \delta_2,
\end{equation}
where $\epsilon$ is an approximated dielectric function, $d_0$ is the thickness of the organic shell, $R_D$ is the radius of the quantum dot core, $V_{\text{cell}}$ is the cell volume of the organic crystal lattice, and $\delta_2$ is a polar-angle dependent, non-uniformity factor Engelmann et al. includes for the shell thickness.  

The problem with this solution in terms of application is that one must carefully choose effective dielectric constants for the model prior to extrapolating data for suitable semiconductor materials. Rather, one must either choose a \textit{good} effective dielectric constant for the system, then figure out the heterostructure makeup like Bohr radius, or vice-versa. Attacking the problem in this manner would be incredibly inefficient and could have wrong assumptions.

We define $\epsilon$ using an internal homogenization effective medium theory (IH EMT). The particular function used is
\begin{equation}
\label{ihemt}
\frac{\epsilon - 1}{\epsilon + 2} = \frac{(\epsilon_s - 1)(\epsilon_c + 2\epsilon_s) + f(\epsilon_c - \epsilon_s)(1 + 2\epsilon_s)}{(\epsilon_s + 2)(\epsilon_c + 2\epsilon_s) + 2f(\epsilon_s - 1)(\epsilon_c -\epsilon_s)},
\end{equation}
where $f = \big(\frac{R_{c}}{R_{s}}\big)^3$, the fraction of the overall sphere volume that the dot core occupies, i.e. the \textit{filling factor}. In this process, the polarizability of the equivalent sphere is equated to that of a core-shell in the electrostatic approximation. Note that \textbf{(\ref{ihemt})} is independent of the particle size and is only valid in the regime for which the particle size is much smaller than the incident wavelength \cite{gutierrez}. We solve \textbf{(\ref{ihemt})} for the effective dielectric constant and further express the solution in terms of a new quantity $\eta$, the ratio of the inorganic core dielectric constant to the organic shell-material dielectric constant
\begin{equation}
\label{ModifiedIHEMT}
\epsilon=\epsilon_s \frac{\eta(1+2f)+2(1-f)}{\eta(1-f)+(2+f)}.
\end{equation}

As one can see, we now have an equation that allows us to analyze the hybridization parameter in terms of the core material, if we choose. This is more desirable for experimentalists because inorganic material have been studied in far greater detail and possess more predictable electronic characteristics. Ideally, we would like to reduce the number of parameters in the mixing coefficient to such an extent that heterostructures could be easily determined using very few semiconductor properties and have their optoelectrical properties closely predicted.

This hybridization parameter, expressed in terms of the filling factor, allows us to examine the system in terms of a ratio relating the core radius and shell thickness,
\begin{equation}
\label{ModifiedEngelmann2}
\Gamma = - \frac{32 \eta [\eta(1-f)+(2+f)] \mu_W \mu_F}{15 \epsilon_c [\eta(1+2f)+2(1-f)]} \delta_2 \sqrt{\frac{\pi f}{2 a_{B}^3 a_{L'}^3 {(1 - \sqrt[3]{f})}^3}}.
\end{equation}
We have also taken the liberty of writing the unit cell volume of the organic lattice as an \textit{isometrized} lattice constant, simply defined as $a_{L'} \equiv \sqrt[3]{V_{\text{cell}}}$. For example, reported lattice parameters for perylene suggest a unit cell volume of around $1230 \, \text{\AA}$, which translates to an \textit{isometrized} latttice constant of around $10.7 \, \text{\AA}$ \cite{fedorov}. The reason for this is to further help us estimate other properties of the heterostructures without compromising earlier work.

The existence of the Wannier-Mott exciton in the dot will restrict our core sizes to those that foster weak-confinement conditions in addition to nonlinear susceptibilities. As detailed in \cite{engelmann},  desired optoelectrical characteristics of the core-shell heterostructure depends significantly on a large, \textit{but not too large}, and small, \textit{but not too small}, core radius. We must determine a regime in which the semiconductor excitons interact, allowing for nonlinear optical properties, and the weak-confinement limit is obeyed \cite{engelmann}.
In a least restrictive domain, we take the lower bound and upper bound for the core radius as $a_B \leq R_{c} \leq R_{\text{max}}$, where $R_{\text{max}}$ is the half-wavelength required for the exciton state transition. As discussed by Steiner, this upper bound should be no larger than $10 \, \text{nm}$ for a certain room temperature heterostructure \cite{steiner}.

For $\epsilon_c$, using calculated Bohr radii and generally accepted experimental dielectric constants of various common semiconductor materials, we obtained a statistical fit relating exciton Bohr radii and dielectric constants 
\begin{equation}
\label{bohrRadiustoDielectricConstant}
\epsilon_c \approx 2.28311 \ln{\bigg[\frac{a_B \times 10^{10}}{0.628}\bigg]}.
\end{equation}
The fit above, \textbf{(\ref{bohrRadiustoDielectricConstant})}, calculated from the data in \textbf{Fig. \ref{bRadiusDielectricPlot}} below, is a characteristic of this model that should be adjusted as more experimental and theoretical data become available. More accurate curve-fitting will enhance the model and allow for more predictable heterostructures.
\begin{figure}[htbp]
  \centering
   \includegraphics[width=0.5\textwidth]{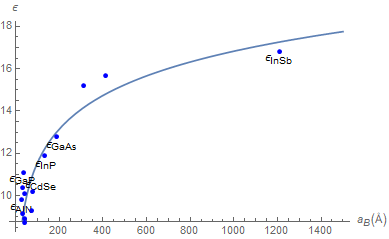}
  \caption{Calculated Bohr radii and dielectric constants for various common, inorganic semiconductor materials with logarithmic curve-fitting.}
\label{bRadiusDielectricPlot}
\end{figure}
With this expression of $\epsilon_c$, the hybridization parameter can be expressed in terms of the materials' dipole moments, in addition to the ratio of their dielectric constants, the filling factor, the core Bohr radius and the organic material unit cell volume.

\section{\label{Polariton}The Organic-Inorganic Hybrid Polariton}
We study the formation of organic-inorganic hybrid excitonic polariton by excitation of an organic-inorganic heterostructure in a microcavity by an optical wave. As noted above, the organic-inorganic hybrid exciton has a large exciton radius from the inorganic component and a very large oscillator strength from the organic component, all in addition to an increased optical nonlinearity by 2 orders. At resonace, the hybridization will encourage the formation of the polariton at high temperatures.

We consider a semiconductor-organic system consisting of the Wannier-Mott exciton, the Frenkel exciton, and the photon with interaction between them
 \begin{equation}
\label{InteractionPolaritonHamiltonian}
\begin{aligned}
H_{\pi} &=H_{F}+H_{W}+H_{F-W}+H_{\gamma}+H_{F-\gamma}+H_{W-\gamma} \\
&= \sum_{\mathbf{k}}{}E_{\mathbf{k}}^{F}\hat a_{\mathbf{k}}^{\dagger}\hat a_{\mathbf{k}}+E_{\mathbf{k}}^{W}\hat b_{\mathbf{k}}^{\dagger}\hat b_{\mathbf{k}}+\Gamma_{\mathbf{k}}(\hat a_{\mathbf{k}}^{\dagger}\hat b_{\mathbf{k}}+\hat b_{\mathbf{k}}^{\dagger}\hat a_{\mathbf{k}})+E_{\mathbf{k}}^{\gamma}\hat c_{\mathbf{k}}^{\dagger}\hat c_{\mathbf{k}}+\zeta_{\mathbf{k}}(\hat a_{\mathbf{k}}^{\dagger}\hat c_{\mathbf{k}}+\hat c_{\mathbf{k}}^{\dagger}\hat a_{\mathbf{k}})+\xi_{\mathbf{k}}(\hat b_{\mathbf{k}}^{\dagger}\hat c_{\mathbf{k}}+\hat c_{\mathbf{k}}^{\dagger}\hat b_{\mathbf{k}}),
\end{aligned}
\end{equation}
where $H_{\gamma}$ is the pure photon Hamiltonian, $H_{F-\gamma}$ and $H_{W-\gamma}$ are the interaction Hamiltonians describing the coupling between the Frenkel exciton and photon with coupling constant $\zeta_{\mathbf{k}}$ and the Wannier-Mott exciton and photon with coupling constant $\xi_{\mathbf{k}}$ respectively.

From the Hamiltonian (12) a pure hybrid polariton system has been obtained
\begin{equation}
\label{HPolaritonHamiltonian}
H_{\pi}=\sum_{\mathbf{k}}{E_{\mathbf{k}}^{\pi}\hat \pi_{\mathbf{k}}^{\dagger}\hat \pi_{\mathbf{k}}}.
\end{equation}

using the unitary transformation $U$,
\begin{equation}
\label{HPolaritonUnitaryTransformation}
\begin{pmatrix}
\hat a_{\mathbf{k}} \\
\hat b_{\mathbf{k}} \\
\hat c_{\mathbf{k}}
\end{pmatrix}
= U
\begin{pmatrix}
\hat \pi_{\mathbf{k}1} \\
\hat \pi_{\mathbf{k}2} \\
\hat \pi_{\mathbf{k}3}
\end{pmatrix}
\end{equation}
which has the characteristic equation
\begin{equation}
\label{HPolaritonCharacteristicEquation}
(E_{{\mathbf{k}}j}^\pi - E_{\mathbf{k}}^\gamma)(E_{{\mathbf{k}}j}^\pi - E_{\mathbf{k}}^F)(E_{{\mathbf{k}}j}^\pi - E_{\mathbf{k}}^W) - [\Gamma_{\mathbf{k}}^2(E_{{\mathbf{k}}j}^\pi - E_{\mathbf{k}}^\gamma) + \zeta_{\mathbf{k}}^2(E_{{\mathbf{k}}j}^\pi - E_{\mathbf{k}}^W) + \xi_{\mathbf{k}}^2(E_{{\mathbf{k}}j}^\pi - E_{\mathbf{k}}^F) + 2\Gamma \zeta \xi] = 0.
\end{equation}
with solution as three branches for energies of the hybrid polariton state.

\begin{figure}[htbp]
  \centering
   \includegraphics[width=0.65\textwidth]{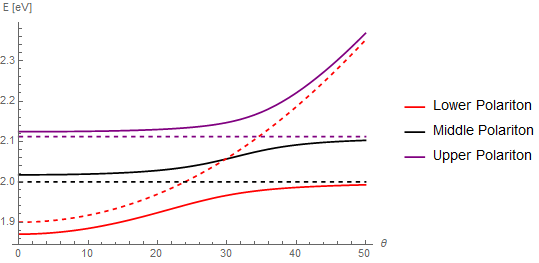}
  \caption{Relationship between the angle of incidence $\theta$ and the polariton energy.}
\label{hPolaritonDispersionCurves}
\end{figure}

The energy of the microcavity photon with refractive index $n$ is $E_{\mathbf{k}}^\gamma = E_{0}^\gamma \Big[1-\frac{\sin^2 \theta}{n^2}\Big]^{-1/2}$, and the Frenkel exciton and Wannier-Mott exciton energies are taken as $E_{\mathbf{k}}^F = E_{0}^W + \delta$ and $E_{\mathbf{k}}^W = E_{0}^W$ respectively. The eigenenergies of the hybrid polariton can be plotted as a function of $\theta$, as shown in \textbf{Fig. \ref{hPolaritonDispersionCurves}}, where $\Gamma_{\mathbf{k}}$ in this case is considered vanishingly small.

\begin{figure}[htbp]
\centering
\begin{subfigure}[b]{.5\textwidth}
  \centering
  \includegraphics[width=0.9\linewidth]{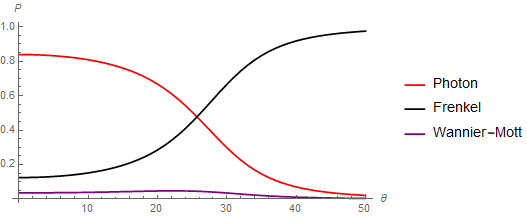}
  \caption{Lower branch mixing coefficients.}
  \label{hPolaritonLowerMixingCurves}
\end{subfigure}%
\begin{subfigure}[b]{.5\textwidth}
  \centering
  \includegraphics[width=0.9\linewidth]{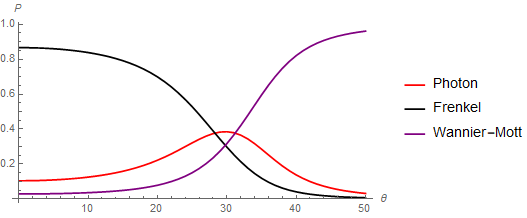}
  \caption{Middle branch mixing coefficients.}
  \label{hPolaritonMiddleMixingCurves}
\end{subfigure}%

\begin{subfigure}[b]{.5\textwidth}
  \centering
  \includegraphics[width=0.9\linewidth]{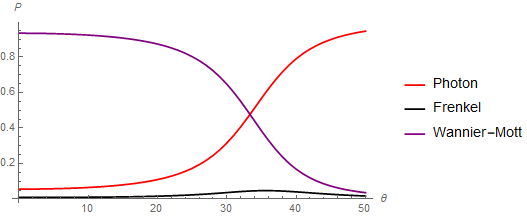}
  \caption{Upper branch mixing coefficients.}
  \label{hPolaritonUpperMixingCurves}
\end{subfigure}
\caption{Relationship between the angle of incidence $\theta$ and the branch mixing coefficients.}
\label{polaritonbranches}
\end{figure}

Additionally, the mixing coefficients for the lower, middle, and upper hybrid polariton branches are plotted in \textbf{Fig. \ref{hPolaritonLowerMixingCurves}}, \textbf{Fig. \ref{hPolaritonMiddleMixingCurves}}, and \textbf{Fig. \ref{hPolaritonUpperMixingCurves}} respectively.
Note the change of the mixing coefficients, or the weights of different excitonic species in the different branches of the mixed polariton state. This suggests choosing the branch suitable for the application, depending on the preferred properties of a specific excitonic species.

\section{\label{Results}Numerical Results and Discussion}
\begin{figure}[htbp]
\centering
\begin{subfigure}[b]{.5\textwidth}
  \centering
  \includegraphics[width=0.85\linewidth]{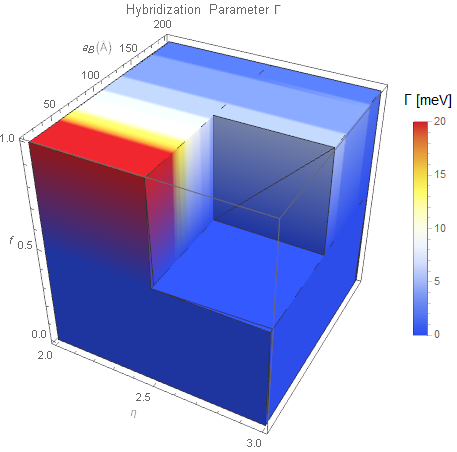}
  \caption{}
  \label{sliceDensityPlotCenterCutBox}
\end{subfigure}%
\begin{subfigure}[b]{.5\textwidth}
  \centering
  \includegraphics[width=0.85\linewidth]{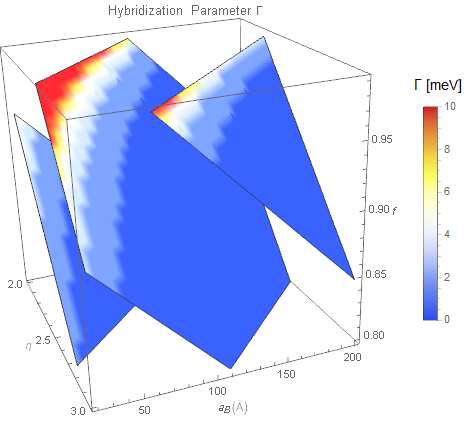}
  \caption{}
  \label{sliceDensityPlotDiagonalStackedPlanes}
\end{subfigure}
\caption{Density plots for the hybridization parameter dependent on the semiconductor exciton Bohr radius, the core-shell filling factor, and the ratio of inorganic to organic dielectric constants.}
\label{hybridizationparameter}
\end{figure}

From the results obtained in \textbf{(\ref{ModifiedEngelmann2})}, a density plot of the hybridization coefficient is presented below, where both dipole moments are valued at $1 \, \text{D}$, a \textit{non-uniformity factor} \cite{engelmann} of $\frac{1}{10}$, and an isometric organic lattice constant of $5 \, \text{\AA}$. \textbf{Fig. \ref{sliceDensityPlotCenterCutBox}} allows us to ascertain heterostructures that exhibit specific exciton hybridization between the organic shell and inorganic core. As an example, one might achieve a hybridization parameter magnitude $\approx 15 \, \text{meV}$ for a core-shell nanoparticle (CSNP) with a Wannier-Mott exciton Bohr radius of around $50 \, \text{\AA}$, an inorganic dielectric constant between $2$ to $3$ times the organic shell material dielectric constant, and a CSNP filling factor greater than about $0.6$.

\begin{figure}[htbp]
  \centering
   \includegraphics[width=0.5\textwidth]{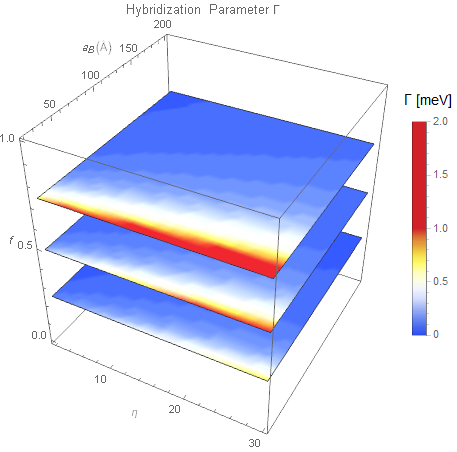}
  \caption{A sliced density plot that illustrates the relatively negligible influence that the dielectric ratio has on the hybridization parameter.}
\label{sliceDensityPlotZStackedPlanes}
\end{figure}

From \textbf{Fig. \ref{sliceDensityPlotCenterCutBox}}, it is still difficult to tell how one achieves a specific interaction value; however, the following plot, \textbf{Fig. \ref{sliceDensityPlotDiagonalStackedPlanes}}, demonstrates the subtle effect the filling factor has on the system. It is interesting, and strange, to note that the dielectric ratio $\eta$ plays a very small role in the hybridization parameter. \textbf{Fig. \ref{sliceDensityPlotZStackedPlanes}} shows just how negligible the parameter is if we consider only values between $2$ to $3$. Given a much larger domain, one which at present may be overly estimating our choices for organic materials, we can see that the dielectric ratio becomes significant just outside of the bound we consider, i.e. $\eta \gtrsim 3$, and only to a factor of around $\frac{1}{10}$.

It is somewhat of a subtlety, and worth mentioning, how influential the non-uniformity factor is in the hybridization between the two exciton species. A degree of \textit{roughness} at the interface of the core and shell could potentially be exploited to repress any shortcoming with other parameters of the system. We take special notice of this because altering any, say, \textit{lengths} in the parameter is more tedious to track. Stated better perhaps, once we choose our heterostructure materials, we want to stick with them. Thus, we look to tailor other variables to enhance a particular attribute of the system. This would include the interface \textit{roughness} and the average thickness of the shell.

\begin{figure}[htbp]
\centering
\begin{subfigure}[b]{.5\textwidth}
  \centering
  \includegraphics[width=0.9\linewidth]{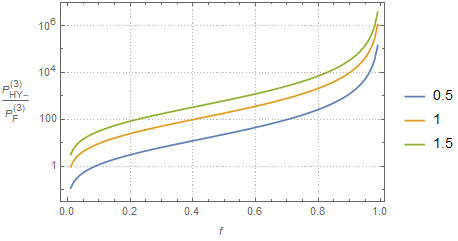}
  \caption{}
  \label{MagnitudeOpticalNonlinearityLogPlot}
\end{subfigure}%
\begin{subfigure}[b]{.5\textwidth}
  \centering
  \includegraphics[width=0.9\linewidth]{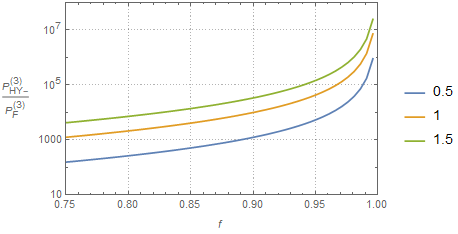}
  \caption{}
  \label{MagnitudeOpticalNonlinearityLogPlotZoom}
\end{subfigure}
\caption{A log-plot for the rough orders of magnitude of third-order nonlinearity ratios dependent on the CSNP filling factor. Each curve is a different average shell thickness to organic lattice constant ratio. \textbf{\subref{MagnitudeOpticalNonlinearityLogPlotZoom})} is the desired filling factor region.}
\label{nonlinearity}
\end{figure}

The optical nonlinear susceptibility, $\chi$, for the hybrid exciton at resonance, again detailed in \cite{engelmann}, as a ratio with respect to the Frenkel species of exciton has orders of magnitude plotted in \textbf{Fig. \ref{nonlinearity}} against the filling factor for the CSNP heterostructure. The three different plots refer to three various ratios of the average shell thickness, $d_0$, to the \textit{isometrized} organic lattice constant, $a_{L'}$, 
\begin{equation}
\label{thirdordernonlinearityratio}
\frac{P^{(3)}_{HY-}}{P^{(3)}_{F}} \sim 4 \pi \Bigg( \frac{d_0}{a_{L'}} \Bigg)^3 \Bigg( \frac{\sqrt[3]{f}}{1-\sqrt[3]{f}} \Bigg)^2.
\end{equation}
It is obvious that the hybridization is a critical ingredient in enhancing the third-order nonlinearity.

\begin{figure}[htbp]
\centering
\begin{subfigure}[b]{.5\textwidth}
  \centering
  \includegraphics[width=0.9\linewidth]{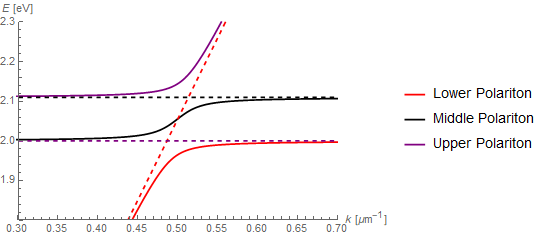}
  \caption{}
  \label{HPolaritonDispersionGamma1Mathematica}
\end{subfigure}%
\begin{subfigure}[b]{.5\textwidth}
  \centering
  \includegraphics[width=0.9\linewidth]{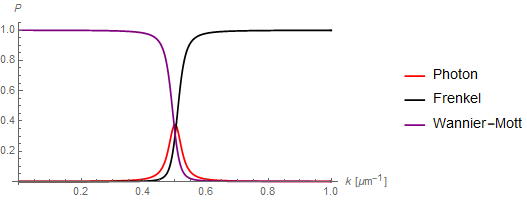}
  \caption{}
  \label{HPolaritonMiddleMixingGamma1Mathematica}
\end{subfigure}%

\begin{subfigure}[b]{.5\textwidth}
  \centering
  \includegraphics[width=0.9\linewidth]{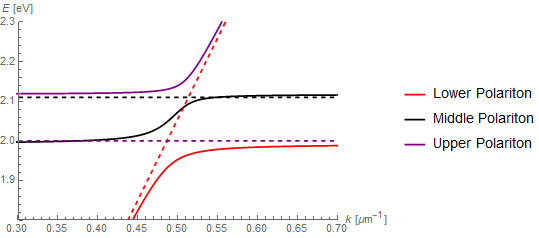}
  \caption{}
  \label{HPolaritonDispersionGamma30Mathematica}
\end{subfigure}%
\begin{subfigure}[b]{.5\textwidth}
  \centering
  \includegraphics[width=0.9\linewidth]{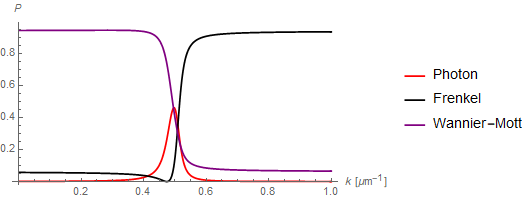}
  \caption{}
  \label{HPolaritonMiddleMixingGamma30Mathematica}
\end{subfigure}
\caption{The dispersion and middle branch mixing curves for the hybrid polariton in the case of a CdSe core. \textbf{\subref{HPolaritonDispersionGamma1Mathematica})} and \textbf{\subref{HPolaritonMiddleMixingGamma1Mathematica})} illustrate the case when the hybridization parameter is $1 \, \text{meV}$, while \textbf{\subref{HPolaritonDispersionGamma30Mathematica})} and \textbf{\subref{HPolaritonMiddleMixingGamma30Mathematica})} depict the case when the hybridization parameter is $30 \, \text{meV}$.}
\label{hybridpolariton}
\end{figure}

For two cases of the hybrid polariton, the eigenenergies and middle branch mixing coefficients are plotted alongside one another without the consideration of a microcavity, with exciton energies as those approximated using the Brus equation, i.e. using the effective mass approximation, and the photon energy as $E_{\mathbf{k}} = c \hbar k$. As shown in \textbf{Fig. \ref{hybridpolariton}}, the initial Frenkel exciton energy is $2.11 \, \text{eV}$, initial Wannier-Mott exciton energy is $2 \, \text{eV}$, and the heterostructures are such that they yield hybridization values of $1 \, \text{meV}$ and $30 \, \text{meV}$. Both exciton-photon interaction potentials are valued as $50 \, \text{meV}$. 

\section{\label{Conclusion}Conclusion}
General models for both organic-inorganic excitons and polaritons are obtained, and an analytical discussion on the relationship between certain parameters and interactions is provided for the purpose of enhancing the hybridization between the materials.


\bibliographystyle{unsrt}  
\bibliography{references}  


\end{document}